\documentclass[12pt]{article}
\usepackage[margin=0.7in]{geometry}
\usepackage{float}
\usepackage{amsmath}
\usepackage{amssymb}
\usepackage{graphicx}
\usepackage{cite}
\usepackage{color}
\usepackage{dcolumn}
\usepackage{bm}
\usepackage{newfloat}
\DeclareFloatingEnvironment[name={SI Figure}]{suppfigure}
\DeclareFloatingEnvironment[name={SI equation}]{suppequation}
\DeclareFloatingEnvironment[name={SI Table}]{supptable}

\usepackage{amsmath}
\usepackage{graphicx}
\usepackage{dcolumn}
\usepackage{bm}
\usepackage{amsmath}
\usepackage{amssymb}
\usepackage{amsfonts}
\usepackage{graphicx}
\usepackage{dcolumn}
\usepackage{bm}
\usepackage{sidecap}
\usepackage {caption}
\usepackage{float}
\usepackage{parskip}

\usepackage{hyperref}
\usepackage{longtable}
\usepackage{array}
\usepackage{array,booktabs,arydshln,xcolor}

\usepackage{booktabs}

\usepackage{graphics}
\usepackage{xcolor,colortbl}

\definecolor{Gray}{gray}{0.2}
\definecolor{LightCyan}{rgb}{0.2,0.2,1}

\newcolumntype{a}{>{\columncolor{Gray}}c}
\newcolumntype{b}{>{\columncolor{white}}c}

\begin{document}

\begin{flushleft}
{\large
\textbf{Dissortativity and duplications in Oral cancer}
}
\\
Pramod Shinde$^{1}$, Alok Yadav$^{2}$, Aparna Rai$^{1}$, and Sarika Jalan$^{1,2,*}$
\\ 
\it ${^1}$ Centre for Biosciences and Biomedical Engineering, Indian Institute of Technology Indore, Indore 452017, India\\
\it ${^2}$ Complex Systems Lab, Discipline of Physics, Indian Institute of Technology Indore, Indore 452017, India\\

$\ast$ E-mail: sarika@iiti.ac.in
\end{flushleft}

\begin{abstract}
More than 300,000 new cases worldwide are being diagnosed with oral cancer annually. 
Complexity of oral cancer renders designing drug targets very difficult. 
We analyse protein-protein interaction network for the normal and oral cancer tissue and detect crucial changes in the structural properties of the networks in terms of the interactions of the hub proteins and the degree-degree correlations. 
Further analysis of the spectra of both the networks, while exhibiting universal statistical behavior, manifest distinction in terms of the zero degeneracy, providing insight to the complexity of the underlying system.
\end{abstract}

{\color{red}} Oral cancer refers to a subgroup of head and neck malignancies that develop at the lips, tongue, salivary glands, gingiva, floor of the mouth, oropharynx, buccal surfaces and other intra-oral locations  according to the international classification of diseases \cite{WHO}. 
More than 300,000 new cases worldwide are being diagnosed with Oral squamous cell carcinoma (OSCC) annually \cite{CRUK}. 
WHO estimates it particularly to be the eighth most common cancer worldwide \cite{Coelho} and it is of significant public health importance to developing countries such as in Indian subcontinent where it ranks among the top three types of cancer and accounts for over 30\% of all cancers \cite{Sankaranarayanan,KhanZU,DikshitR}. 
Despite the growing propensity for oral cancer, majority of the research works have focused only on breast cancer \cite{BreastCancer}, colon cancer \cite{ColonCancer} and lymphomas \cite{lymphoma,cancers}.
Hence, the identification of genetic changes specific to oral cancer is very crucial because it provides an opportunity to use this information for the development of drugs to treat this disease \cite{Calixto}. 
Availability of huge amount of proteomic data \cite{proteomicdata} and information on the highly interlinked internal organization of the cell metabolism, signal transduction, transport etc. \cite{InterlinkedInfo}, provides us a scope to identify and understand, the principles that govern the behaviour of cells. We analyse the proteomic data using the network theory framework \cite{BarabasiReview,BarabasiNat} which is holistic approach that enables us to capture the intrinsic properties of the underlying system in a cost-effective and time-efficient manner.

Previous attempts to understand various diseases under the network theory reveal that various types of cancers are interlinked through some pathways which are altered in different diseases \cite{GohKI}.
This study explores the networks of the normal oral tissue and its diseasome, by characterizing their structural and spectral properties.
This approach yields a comprehensive picture of the patterns and principles governing oral cancer which otherwise would not be apparent from the study of individual proteins \cite{individual_pro}.
We find significant variations between the structural and spectral properties of the Normal and Cancer network.
The change in the degree-degree correlation provides quantitative information about the consequences of altered cell behaviour from the normal to cancer state.
Further, the node duplication presents a clue to the abrupt transformation of the normal cells to the cancer cells which can be further used to predict potential therapeutics.

\section*{Results}
\begin{table*}\centering
\begin{center}
\caption{Various properties of PPI networks. The total number of proteins $N$ collected using various databases, number of proteins in the largest connected cluster $N_{LCC}$ and connections $N_{C}$, average degree $\langle k \rangle$, average clustering coefficient  $\langle CC \rangle$, degree-degree correlation ($r$), zero eigen values ($\lambda_{0}$) and duplicate nodes ($D_{0}$).}
\begin{tabular}{c c c c c c c c c c}    \hline \specialrule{1pt}{0pt}{0pt}
{\bf Network}  	& {\bf $N$}  &{\bf $N_{LCC}$} & {\bf $N_{C}$} & {\bf $\langle k \rangle$} & {\bf $D$}  & {\bf $\langle CC \rangle$}    &   {\bf $r$} & {\bf $\lambda_{0}$} & {\bf $D_{0}$}\\ \hline \specialrule{1pt}{0pt}{0pt}

Normal		& 2464  & 2105	& 21746	&  21 	& 9     &  0.308 	& 0.180       &60      & 56 \\ 
Cancer		& 1544   & 1542	& 26794 	&  35	& 7     	&  0.352 	& -0.031      &15      & 13\\
NNC	     	& 1396  & 1356	& 10050	& 15 	& 10  	&  0.288   & 0.389       & 34     & 28\\ 
CNC		    & 833  & 802	& 7802  &  20   &  7  	&  0.365 	& -0.056	      &25      & 22\\ 
Common		& 700  & 696	& 7292	& 21	    & 6   	&  0.346 	& 0.050       & 2       & 1\\ \hline \specialrule{1pt}{0pt}{0pt}
\end{tabular}
\begin{flushleft}
\end{flushleft}
\label{network_parameter}
\end{center}
\end{table*}

The Normal dataset contains 2105 nodes (proteins) and 21746 connections (interactions) while the Cancer dataset consists of 1542 nodes and  26794 connections (Table \ref{network_parameter}). 
We only consider the largest connected component ($LCC$) for our analysis.
The Cancer network consists of less number of proteins than the Normal one,
{in part because the Cancer network focuses only on proteins directly implicated in the formation of cancerous tumours \cite{Watts}.}
To address the large-scale structural organization of PPI networks, we first examine the structural properties of these networks.
The Cancer network displays a relatively smaller diameter as compared to the normal one.
We remark that one can have very different values of the diameter for networks having same values of $N$ and $N_{C}$ (number of connections).
Similarly, different values of $N$ and $N_C$ can yield same diameter as it is the manner in which connections are distributed in the network that affects the diameter.
For instance, a regular lattice with size $N$ and average degree $\langle k \rangle$ has diameter which scales as $\sim N / \langle k \rangle$, whereas a random network with the same size and the average degree has the diameter $\sim (ln  N) / (ln \langle k \rangle)$ \cite{Watts}. 
What follows that the smaller diameter observed here for the Cancer network may be due to both the smaller size as well as large average degree.
Since a small diameter facilitates fast communications \cite{Watts}, this may be causing the faster signalling resulting in cancer.
It is reported that intrinsic disorder in the cancer proteins facilitates the adaptability of faster signalling in the Cancer network \cite{Signal}.
Furthermore, we find that the average degree is higher in the Cancer network, exhibiting relatively higher order connections than the Normal network
which suggests that the cancer proteins have more tendency to interact with each other.
The other structural properties, namely the diameter and clustering coefficient behave similar to those of many other biological networks, i.e. small diameter close to the corresponding random networks and high clustering coefficient much larger than the corresponding random networks \cite{BarabasiNat,Barabasi,Clauset}.

We find that there is a significant increase in the degree of the hub proteins in the Cancer network as compared to that of the Normal network, particularly for common hub proteins, namely Ubiquitin and ALB (Table \ref{HubProteins}).
Note that the change in the interactions are arising due to deletion or addition of new nodes in the Cancer network, as interactions are taken from STRING database for both the Cancer and Normal datasets.
These hub proteins, on an average, have twice as many interaction partners in the Cancer as compared to the Normal PPI network.
Since the Cancer network is almost twice dense as compared to the Normal network, the observation that the hub proteins in the Cancer have almost twice the degree as compared to much high degree of the hub proteins in the Normal is not surprising.
Both the networks have high degree nodes involved in almost 50\% of the interactions indicating that these connected proteins have significant role in a network.
As it has been reported that high degree proteins tend to share certain relevant functional information in their respective pathways \cite{Hsing}, we study these hub proteins in both the Normal and the Cancer networks. 
In order to understand the impact of increased interactions of the hub proteins in the Cancer, we explore the functional and structural aspects of these proteins using protein data bank \cite{pdb}.
We find that these proteins have large molecular organizations with structural and functional domains that facilitates maximum interactions with other proteins, and are perceived for their implicit roles such as functional characterization, pathway analysis, structural alterations etc in cancer molecular pathways \cite{Bio2006}.
\begin{table}[ht]   
\begin{center}
\caption{ Biological functions of Hub proteins. $k$ denotes degree of protein in the Normal and Cancer network.}
\begin{tabular} {p{2cm} p{0.7cm} | p{2cm} p{0.7cm}}  
\hline \specialrule{1pt}{0pt}{0pt}
\multicolumn{2}{c|}{\bf Normal} & \multicolumn{2}{c} {\bf Cancer} \\ \hline \specialrule{1pt}{0pt}{0pt}
Hub Protein 	& $k$  & Hub Protein  & $k$   \\ \hline \specialrule{1pt}{0pt}{0pt}
ALB		    & 289   &   Ubiquitin  &  696   \\ 
Ubiquitin	& 245	&   TSPO 	&   438      \\ 
            &    	& 	ALB	  & 411        \\ \hline \specialrule{1pt}{0pt}{0pt}
\end{tabular}
\label{HubProteins}
\end{center}
\end{table}
The first hub protein, Ubiquitin is known to play an important role in protein processing and recognition.
The deregulation, mutation and over-expression of ubuquitin is reported in the cancer development, consequence to increase interactions of ubiquitin in the central cellular pathways that control signalling, cell cycle, endocytosis, cell death as well as interactions between pathways of the tumour and its surrounding tissue \cite{Kessler,Haglund}.
The increased interactions of the second hub protein, ALB, in cancer is not surprising as concentration of total ALB is known to associate with the development oral cancer \cite{Shpitzer,Nagler}. 
The next hub protein in cancer, TSPO, is known to be highly regulated in normal oral tissues and a significant increase in the number of TSPO interactions in the Cancer network is supported by the fact that it is known to be over-expressed in a highly aggressive oral cancer tumour \cite{Mukherjee}.
Although, the detailed mechanism of TSPO affecting the oral cancer is far from being clear, it is reported to be critical in the cancer development and progression of cancer \cite{Nagler,Mukherjee}.
The increased localization of hub proteins in cancer mucosa explains their role in the formation of the proteome backbone since majority of molecular interactions in the network are channelized by them.

The salient role of change in the interactions of hub proteins in cancer is revealed through the degree-degree correlations, and spectral behaviour.
The Normal network turns out to be assortative whereas the Cancer network indicates dissortative degree-degree correlation with the value of $r$ being negative and close to zero (Table {\ref{network_parameter}).
In order to understand the difference between the degree-degree correlation behaviour of the Normal and Cancer networks, we investigate the NNC, CNC, and Common networks as degree-degree correlation of these networks contribute to those of the Normal and Cancer network.
After comparison, we find that though the NNC and Common networks have positive degree-degree correlations, the value of  $r$ in the NNC network is much higher than that of the entire Normal network (Table \ref{network_parameter}). 
It clearly suggests that the assortativity in the Normal PPI network is solely due to the interactions among the proteins which are functioning in the normal oral tissues only.
The Cancer network exhibiting a negative degree-degree correlation indicates that high degree proteins tend to interact with low degree proteins.
The CNC being the dissortative network (Table \ref{network_parameter}) which reflects that the dissortativity in the Cancer network arises mainly due to the interactions among proteins uniquely involved in the Cancer PPI network.

\begin{table*}\centering
\begin{center}
\caption{ Various structural and spectral properties of the configuration models of the Normal and Cancer PPI networks. We take degree sequence ($\{k\}$ = ${k_1, k_2, . . . , k_N}$) of the Normal and Cancer PPI networks to construct corresponding configuration model. A configuration model generates a random model network with exactly given degree sequence. All quantities were averaged over 20 realizations of the configuration networks.}
\begin{tabular} {c c c c c c c}    
\hline \specialrule{1pt}{0pt}{0pt}
{\bf Network}  	& {\bf $N_{LCC}$}  & {\bf $D$}  & {\bf $\langle CC \rangle$}   &   {\bf $r$} & {\bf $\lambda_{0}$} & {\bf $D_{0}$}\\ \hline \specialrule{1pt}{0pt}{0pt}
Normal		& 2105     	& 6	   & 0.064 $\mp$	0.002 	& -0.03 $\mp$ 0.02  & 39 $\mp$ 7   & 21 $\mp$ 5 \\ 
Cancer		& 1542	    & 5    & 0.154 $\mp$ 0.002    & -0.09 $\mp$ 0.05 & 5 $\mp$ 2   & 4 $\mp$ 2 \\ \hline \specialrule{1pt}{0pt}{0pt}
\end{tabular}
\label{network_parameter2}
\end{center}
\end{table*}

What follows that there is a remarkable alteration in the interaction patterns of the Cancer network which is captured by the change from the positive degree-degree correlation in the Normal network to negative degree-degree correlation in the Cancer network.
This alteration in the degree-degree correlation is also supported by the fact that there is a fast cancer progression leading to a rapid accumulation of genes via neo- and sub-functionalization events as well as by the loss of unnecessary genes \cite{Force}, which may result in the alteration of all the properties in the Cancer network.
In order to see whether this change in the degree-degree correlation from the Normal to Cancer is due to the change in the degree sequence, particularly due to the change in the degrees of hubs which we have already discussed to be significantly increased in the Cancer network, we compare all of these networks with their corresponding random control networks.
The method of construction of these configuration networks has been discussed in Material and Methods section.
Various structural properties of corresponding random graphs are listed in Table \ref{network_parameter2}. 
The diameter as well as average clustering coefficient of the corresponding random networks are small as expected \cite{Conf}.
Both the corresponding random networks turn out to be disassortative. 
Although the Normal network is assortative, the configuration networks with the same degree sequence may still differ significantly in various network features interaction between pair of nodes are random here.
This behaviour is in line with the finding that less assortative (more dissortative) a network is more will be the randomness in the underlying system \cite{Alok}}.
Both the Cancer PPI network and its corresponding random network, displaying the negative degree-degree correlation, suggest that the Cancer network is more random as compared to the Normal network.
This further indicates that while randomness is an essential ingredient, contributing to robustness of a system \cite{Kitano}, then too much randomness is detrimental.

\begin{figure}
\centering
\includegraphics[width=0.6\columnwidth]{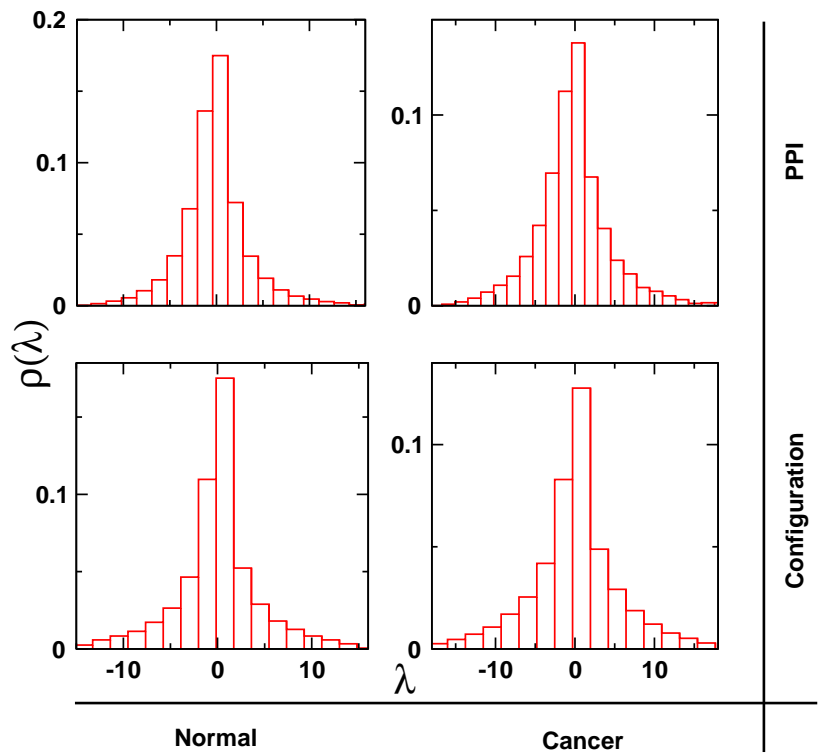}
\caption{(Color online) Eigenvalue distribution ($\rho(\lambda)$) for the Normal and Cancer network datasets of both the PPI and the corresponding configuration model networks.}
\label{Fig_EigenDist}
\end{figure}

So far we have focused on various structural aspects of both the networks in order to understand the distinguishable structural features of the Cancer network from the Normal.
In order to further get insight to the complexity of both the networks, we analyse spectra of these networks.
The eigenvalue statistics of both the PPI networks reflect a triangular shape (Fig. \ref{Fig_EigenDist}), owing to the scale free network behaviour \cite{BarabasiReview}.
Further, spectra for both the PPI networks exhibit a high degeneracy at the zero eigenvalue (Table \ref{network_parameter}).
This high degeneracy at zero eigenvalue is not surprising as many biological networks have been shown to exhibit the similar behaviour \cite{BarabasiReview,Alok}. 
What is captivating that despite similar overall spectral properties, the height of the peak at zero eigenvalue differs considerably in both the networks.
The question lies in whether this difference in the number of zero eigenvalue arises simply due to change in the network size and  average degree or due to the alterations in interaction patterns of the normal and cancer as well as whether this change in zero degeneracy provides any insight to transition from a tissue being normal to cancerous or it is arising due to the considerable increase in the number of interactions of the hub nodes in the cancer as compared to the normal. 
The last column in Table 1 indicates the number of exact duplicates in the networks which contribute to same number of zero eigenvalue. 
Rest of the zero eigenvalues are arising due to existence of the partial duplicates, calculation of which is computational exhaustive \cite{Alok}.
{We provide exact relation between $\lambda_{0}$ and duplicate nodes in Methods and Materials}, as well as the derivation relating to the degree with the duplicate nodes in \nameref{S2_text}.

So a comparison merely based on the size of networks is not appropriate.
A better comparison can be drawn from the corresponding random networks, which have the same size and the average degree ($\langle k \rangle$) as of the Normal and Cancer networks.
Note that instead of a random network of the same size and $\langle k \rangle$, the configuration model is considered, which is a closer random replica of the real networks, as in addition to $N$ and $\langle k \rangle$, it preserves the degree sequence as well.
Hence, in order to gain a better understanding of the origin of zero eigenvalue and hence duplicate nodes, we calculate the spectra of the corresponding configuration model. 
These corresponding random controls also exhibit a high degeneracy at the zero eigenvalue (Fig. \ref{Fig_EigenDist}) but PPI networks have considerably more degeneracy at the zero eigenvalue than those of the corresponding random networks (Table. \ref{network_parameter2}).
This indicates that not only the exact degree sequence but also the contribution of other factors, such as the nature by which the PPI networks have evolved and acquired a pattern may have contributed to the degeneracy at the zero eigenvalue \cite{Aquir}. 
In the following, we delve into understanding the origin behind the occurrence of the degeneracy at the zero eigenvalue.

The occurrence of the zero eigenvalues has a direct relation with the node duplication, as discussed in details in the materials and methods section.
The number of rows following (i), that is number of complete duplications, for both the networks are enlisted in Table 1.
We find that both the Normal and Cancer networks display the property of node duplication.
Further, the Cancer network exhibits fewer duplicate nodes than the Normal network, both in raw count and as a fraction of all the nodes in the network.
The explanation behind this can be given in terms of the possible number of ways in which duplication can happen which is the ratio of the possible number of combinations of duplication possessed by two $k$-degree nodes to the possible number of combinations of random  $k!$ connections of those nodes. 
This is given as $ N! / N^k$ (\nameref{S2_text}), where $N$ is the total number of nodes. 
As $k$ increases, the possible number of ways complete duplication can happen decreases.
Therefore, the existence of larger number of connections among the cancer proteins, as compared to the normal state is expected to result in the less duplication.
The interesting point is that the number of duplication is more than the corresponding random networks (Table \ref{network_parameter2}), importance of which we elaborate 
subsequently.

A high value of duplication in the Normal network is understood as duplication has been emphasized as a consequence of the evolutionary processes over the years \cite{Zhang} conferring robustness to underlying systems \cite{Kitano,Nowak}, but what is interesting is that the Cancer network also displays the existence of complete and partial duplications, and these all duplicated proteins, except one, in the Cancer network are different from those of the Normal network. 
It means that new duplications arise in the course of mutation leading a normal cell to transform into a cancerous cell.
As discussed, we find that the corresponding random networks have significantly less number of duplicate nodes than that of PPI networks. 
It indicates that the occurrences of gene duplications, due to abrupt transformation of the normal cell to cancerous state, have greatly contributed in shaping the global structure of the Cancer PPI network, which is supported by the fact that intrinsic heterogeneity and own decision making strategy of tumours lead to duplicate genes \cite{Frank}.  
Since in the normal networks, duplication occurs in the course of evolution taking place over hundreds of year time scale \cite{Zhang,Richardson}, in the similar line, more 
duplicate of the Cancer network from the corresponding random networks suggest a some kind of evolution, drastically changing the global state of the Normal PPI network.
Furthermore, duplication has been emphasized to confer genomic stability \cite{Bailey}, and is confirmed by the existence of significantly more number of duplication in the Normal network. 
Moreover, the cancer networks are known to be genomically instable  \cite{Creixell}, which is reflected in our analysis through the destruction of duplicate nodes found in the Normal network.

\section*{Discussions}
Most of the empirical studies on large-scale social networks focus on individual node properties in order to unravel various features, such as patterns of homophily between agents \cite{homophily} or topological centrality of social agents \cite{Christakis}. Various real world complex systems encompass multiple types of interactions and it is important to uncover the principles shaping the large-scale organization of complex interactions \cite{Multiplex_new}. The multilayer network approach provides a better framework for investigating the underlying properties of such systems. 
Since the success of a movie is mostly governed by the calibre of the actors involved, apart from the story, direction, budget, etc. the more number of different types of interactions the actors are involved in has been shown to have positive relation with their success. Similarity in association has been shown to propagate positive attributes like cooperation and trust prevailing in the society. Further, benefits derived from association with dissimilar types of nodes as well as sets of close knit recurring motifs have been emphasized upon. This study also sheds light on the importance of every type of node in building up a robust and successful system. Thus, we propagate the idea that an efficient society is build up on positive attributes such as trust and cooperation which arises on working with similar types of people, though dissimilar sets of people can also emerge successful if they show repeated intermingling, eventually emphasizing on the importance every individual of the society, which is again supported by the weak ties analysis. 
The positive emotions demonstrating association betweeen dissimilar types of nodes can be treated as one of the aspects of future investigation and might attract research in disciplines like psychology \cite{Psychology1,Psychology2}.

\section*{Conclusions}

The analysis of the interaction networks of the oral cavity proteome in the normal and cancer states exhibit overall similarity in various structural features with 
small diameter and large clustering coefficient, which are known to be signatures of the underlying complexity of a system \cite{BarabasiReview}. 
The spectral density of both the networks display triangular shape owing to the scale-free nature of the underlying systems, another signature of complexity in real world systems \cite{Farkas_PRE_2001}. 
A deeper structural analysis reveals a striking difference from the normal to the cancerous state. 
The hub proteins of the Cancer network have almost double the interactions as compared to the normal state. 
As already realized for many other diseases \cite{HubProt}, in the present study, the hub proteins reveal their importance particularly in the gene processing and recognition (See Supplementary Material). 
Although these hubs capture the changes in number of interactions from normal to disease state, they fail to provide insight into changes in the overall interaction patterns which is revealed through the degree-degree correlations which turn out to be a quantitative measure of this difference. 
The normal state exhibits the assortative behaviour whereas the cancer state shows a negative degree-degree correlations. 
This indicates a drastic overall change in the interactions from the normal to the disease state leading to a transition from a positive degree-degree correlations to a negative one \cite{NewmanPRE2003,Newman_2004,Assortativity_Plos}. 
More importantly, degree-degree correlation analysis of corresponding configuration network indicates that the Cancer network is more random than the Normal, further suggesting that while randomness is an important ingredient for functionality \cite{Kitano}, and too much randomness is detrimental. 

The spectral analysis of normal network exhibits the degeneracy at the zero eigenvalue which has a direct relation with the duplication, a mechanism adopted by most of the biological systems during evolution \cite{Zhang} to confer genomic stability \cite{Bailey}. 
What is noteworthy that all the node duplications, except one, are destroyed from the normal to the cancer state, suggesting the absence of these nodes involved in genomic instability, one of the plausible reasons behind cancer state.
Additionally, the Cancer network also shows the formation of new duplicates as well as the number of zero eigenvalues is higher than that of the corresponding random model. 
Appearance of new duplicate nodes in cancer provides an insight into the stability of the cancer PPI network in spite of underlying genomic instability which might also be associated with its resistance to drugs \cite{drug_resistance1,drug_resistance2}. Our analysis, while on one hand, provides an insight into the complexity of the disease, a step to understand the cancer, on the other hand, provides the importance of node duplications in underlying systems.

This analysis presented here can be extended further to study complexity of the oral cancer at different stages of the tumour development \cite{TNM} as well as to understand  alterations in personnel traits leading to complexity in patient specific oral tumours \cite{trait}.
These understanding can provide the envelope of the network information which encodes the changes during the progression of the disease.
Further, these altered changes can explain the pathophysiology of the disease which may be used to have new insights into the personalised diagnosis and therapy \cite{tian}.

\section*{Methods}
We prepare an extensive collection of protein interaction data for normal oral tissues and oral cancer from various online databanks.
Only those proteins are taken into account which are reviewed and cited to keep authenticity of the data intact. 
In order to validate this approach, we perform an extensive manual quality check of each protein dataset by using various Bioinformatics databases (See Supplementary Material for details) and Protein Information Resource (PIR) \cite{Mitra}.
The MeSH keywords are used for the literature searching. 

After assimilating the names of proteins for both the normal and oral cancer datasets, their interacting partners are identified and downloaded from STRING $9.1$ database \cite{Wu}. 
We construct five protein-protein interaction (PPI) networks for the normal and cancer tissue to understand the change in their behaviour and patterns in oral cancer. 
The first network (Normal) consists of proteins involved in the normal tissues of oral cavity and surrounding region.
Likewise, second network (Cancer) comprises of proteins which play role or are confirmed to have contributions in formation of oral carcinomas, but not those proteins that indirectly facilitate tumour growth, such as through modifications to metabolic pathways.
Next, we separate out the proteins which are common in the normal oral and cancer tissues for the construction of third network (Common) in order to study the influence of these protein interactions respectively in both the networks as a comparative analysis tool. 
Further, we construct two more networks for two more subtractive behaviour of specific proteins occurring only in the normal (NNC) and cancer (CNC) states, respectively.
Furthermore, we take degree sequence ($\{k\}$ = ${k_1, k_2, . . . , k_N}$) where $N$ is the dimension of the adjacency matrix of the Normal and Cancer PPI networks to construct corresponding configuration model.
A configuration model generates a random model network with exactly given degree sequence \cite{spectra}.
The adjacency matrix or the connection matrix of a network can be written as,
\begin{equation}
A_{\mathrm {ij}} = \begin{cases} 1 ~~\mbox{if } i \sim j \\
0 ~~ \mbox{otherwise} \end{cases}
\label{adj_wei}
\end{equation} 
The most basic structural parameter of a network is the degree of a node ($k_i$), which is defined as a number of edges connected to the node ($k_i=\sum_j A_{ij}$).
The second parameter, the clustering coefficient $(CC)$, is the ratio of the number of interactions a neighbour of a particular node is having and the possible number of connections the neighbours can have among themselves \cite{Newmann} . 
An average clustering coefficient of a network can be written as ${\langle CC \rangle = \frac {1} {n} \sum_{i=1}^{n} CC_i}$.
Further, the network diameter ($D$) is defined as the longest of the shortest paths between all the pair of nodes in a network \cite{Reka}.
\vspace{0.7cm}

\begin{figure}
\centering
\includegraphics[width=0.6\columnwidth]{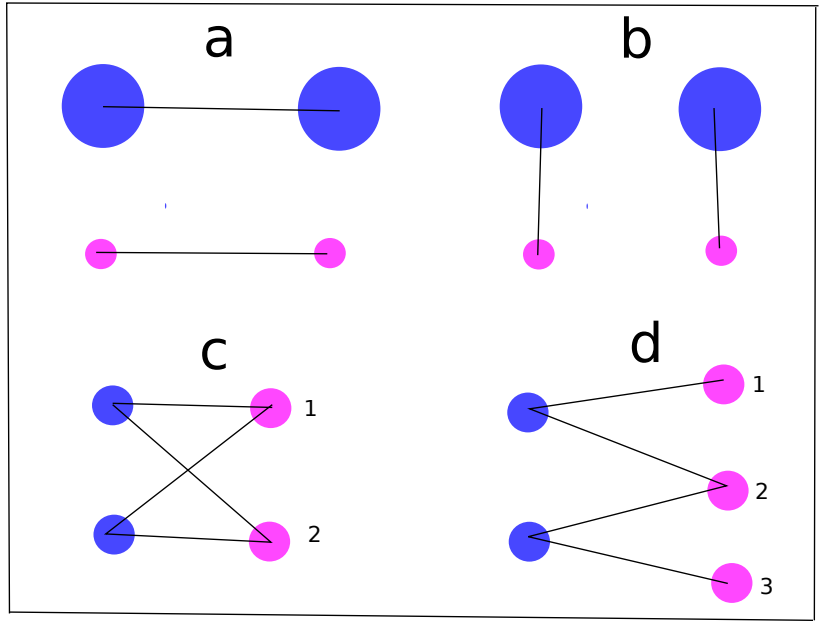}
\caption{(Color online) Schematic description of Assortativity and Node duplication. (a) (1) Assortativity measures the tendency of nodes with the similar numbers of edges to connect. Here, in (a) connection of similar degree nodes,  and (b) connection of dissimilar degree nodes. (2) Few examples of Node duplication (of pink colour nodes) showing (c) Node 1 and 2 have same neighbours and said to be complete duplicates, and (d) Node 2 and 3 are collectively having same neighbours as Node 1 and said to be partial duplicates.}
\label{Fig_Demo}
\end{figure}

Another property of the network which turns out to be crucial in distinguishing the Normal from the Cancer is degree-degree correlation, which measures the tendency of nodes with the similar numbers of edges to connect (Schematic in Fig. \ref{Fig_Demo}).
It can be defined as \cite{Rivera,Alok},
\begin{equation}
r = \frac{[\frac{1}{M} \sum_{i} j_i k_i] - [ \frac{1}{M} \sum_i \frac{1}{2}(j_i + k_i)^2]}
{[ \frac{1}{M} \sum_{i} {\frac{1}{2}} (j_i^2+ k_i^2)] - [ \frac{1}{M} \sum_i \frac{1}{2}(j_i + k_i)^2]}
\label{assortativity}
 \end{equation}
where $j_{i}$ and $k_{i}$ are the degrees of the nodes connected through the $i^{th}$ edge, and $M$ is the number of edges in the network.
The value of $r$ being zero corresponds to a random network where as the negative(positive) value corresponds to dis(assortative) network.

Further, we investigate the role of node duplication by identifying the proteins sharing exactly same neighbours as node duplication is known to be an important  phenomenon occurring during the evolution \cite{Kitano,Ispolatov}. 
The duplicated nodes in a network (Schematic in Fig. \ref{Fig_Demo}) can be identified from corresponding adjacency matrix in the following manner.
When (i) two rows (columns) have exactly same entries, it is termed as the complete row (column) duplication i.e $R_1 = R_2$, 
 (ii) When a combination of rows (columns) have exactly same entries as another combination of rows (columns) then it is termed as the partial duplication of rows (columns), for example  $R_{1} = R_{2} + R_{3}$.
Satisfying any one of the conditions (i) and (ii) lowers the rank of the matrix exactly by one.
In addition, the rank is also lowered if (iii) there is an isolated node.
All these conditions lead to the zero eigenvalues in the matrix spectra.
Since there is no isolated node in the network, ensured in the beginning itself by considering only the largest connected cluster for this analysis,
conditions (i) and (ii) are the only conditions responsible for the occurrence of zero degeneracy.
Counting the number of rows following condition (i) is straightforward, the same is not true for the condition (ii) as there are $\frac {N!} { 2(N-x-y)! }$ (\nameref{S1_Text}) number of ways in which x and y rows can satisfy condition (ii) and hence making the calculations of partial duplications becomes computationally exhaustive \cite{Alok}.
The number of zero eigenvalue provides an exact measure of (i) and (ii) conditions \cite{Van}.
The number of non-zero eigenvalue of a matrix is exactly equal to the rank of the matrix \cite{MatrixA}.
Let us denote eigenvalues of the adjacency matrix by $\lambda_i$, $i = 1, 2, . . . , N$ such that $\lambda_1$ $ < $ $\lambda_2$ $ < $ $\lambda_3$ $ < $ . . .$ <$ $\lambda_N$.

\section*{Appendix}

\subsection*{Appendix A} \label{S1_Text}
{\bf The possible number of ways in which duplication can happen.}\\

\noindent We look for the duplication of $x$ nodes by $y$ nodes in an adjacency matrix of network of $N$ nodes.
Here, we have to look for all the possibilities of corresponding to $x$ and $y$. where $1 < x,y < (N-1)$.
\vspace{3mm} 

\noindent Now, we can choose $x$ nodes out of $N$ nodes in $N \choose x$ number of ways. \vspace{2mm}

\noindent Similarly, we can choose $y$ nodes out of these $(N-x)$ nodes in $ N-x \choose y $ number of ways.\vspace{3mm}
\noindent Therefore, total number of ways in which we can select these $(x ,y)$ pair is \vspace{2mm}

$N \choose x $ $N-x \choose x $ = $\frac{N!}{(N-x-y)! x! y!}$\vspace{3mm}

\noindent As the number of nodes involved are very few, we can neglect $x!$ and $y!$ being order of $\sim 1$ and,

\noindent Since all these selections are symmetric i.e. $AB = BA$, therefore in total \vspace{3mm}

\hspace{14pt}$\sum\limits_{x,y=1}^{N-1}$ \hspace{20pt} $\frac {N!} { 2.(N-x-y)! }$\\ 
{\scriptsize $2\leq (x+y)\leq N$} \vspace{3mm}

\noindent Since $N$ is very large $(\sim 10^3)$ quantity, finding partial duplicate nodes is computationally exhaustive process.

\subsection*{Appendix B} \label{S2_text}
{\bf The probability of two nodes of $k$ degree being duplicated nodes.}\vspace{3mm}

\noindent To explain this, we consider number of ways in which two $k$-degree nodes can select their neighbours excluding themselves is 
$\frac{[(N-2)(N-3)....(N-3-k)]}{k!}^{2}$ $={N-2\choose k}^{2}$ \vspace{2mm}

\noindent and number of ways in which these two $k$-degree nodes select same neighbours \vspace{2mm}

        ${N-2\choose k}{k\choose k} = \frac{[(N-2)(N-3)...(N-3-k)]}{k!}$ \vspace{2mm}

\noindent Therefore, for two nodes of $k$-degree to be duplicated is given by \vspace{2mm}

$\frac{{N-2\choose k} {k\choose k}}{{N-2\choose k} {N-2\choose k}}$ = $\frac{k!}{(N-2)(N-3) . . . (N-3-k)}$ \vspace{2mm}

\noindent If k $\ll$ N, then \vspace{2mm}
$(N-2) \approx N$ ; $(N-3) \approx N$ ;  . . .; $(N-3-k) \approx N$\\

$\hspace{51pt}= \frac{k!}{N^k}$ \vspace{2mm}

\noindent Hence, the probability of two nodes of $k$ degree being duplicated nodes is $\frac{k!}{N^k}$.

\pagebreak

\section*{Acknowledgment}
SJ is grateful to Department of Science and Technology (DST), Government of India and Council of Scientific and Industrial Research,  Government of India project grants EMR/2014/000368 and 25(0205)/12/EMR-II for financial support. 
PS acknowledges DST,  Government of India, for the INSPIRE fellowship (IF150200) as well as the Complex Systems Lab members for timely help and useful discussions.
We thank Dr. Kazuyuki Aihara (University of Tokyo, Japan), Dr. Seyed Hasnain (IIT Delhi, India), and Dr. Syam Prakash Somasekharan Ramachandran Nair (IIT Indore, India) for valuable comments.	

\newpage

\cleardoublepage


\begin{thebibliography}{99}
\bibitem{WHO}
World Health Organization, International Classification of Diseases 10th Revision, Accessed on Jan 20, 2015 (2010).
 
\bibitem{CRUK} Cancer Research UK. Oral cancer incidence statistics, Available from: http://www.cancerresearchuk.org/cancer-info/cancerstats/types/oral/incidence/uk-oral-cancer-incidence-statistics, Accessed on Jan 20, 2015 (2012).

\bibitem{Coelho}
K.R. Coelho, Journal of Cancer Epidemiology. {}{2012} {(2012)}.

\bibitem{Sankaranarayanan}
R. Sankaranarayanan, K. Ramadas, and G. Thomas, {Lancet.} {\bf 9475} {365} {(2005)}.

\bibitem{KhanZU}
Z.U. Khan {Webmed Central CANCER.} {\bf 3(8)} {WMC003626.} {(2012)}.

\bibitem{DikshitR}
R. Dikshit, and P.C. Gupta, {Lancet.} {\bf 9828} {379} {(2005)}.

\bibitem{BreastCancer}
J. Penninger, and D. Schramek, {U.S. Patent Application and references therein.} {\bf 13/825} {655} {(2011)}.

 \bibitem{ColonCancer}
S. Di Franco, P. Mancuso, A. Benfante, M. Spina, F. Iovino, F. Dieli,  and M. Todaro, {Cancers.} {\bf 3(2)} {1957-1974} {(2011)}.

\bibitem{lymphoma}
B. Coiffier, {J Clin Oncol Res.} {\bf 23(26)} {6387-6393} {(2005)}.

 \bibitem{cancers}
National Institute of Cancer.  {http://www.cancer.gov/cancertopics/druginfo/drug-page-index,}{}{} {Accesed on 26th January, 2015}.

\bibitem{Calixto}
G. Calixto, J. Bernegossi, B. Fonseca-Santos, and M. Chorilli {Int. J. Nanomedicine.} {\bf 9} {3719} {(2014)}.

 \bibitem{proteomicdata}
S. Hu, M. Arellano, P. Boontheung, J. Wang, H. Zhou, J. Jiang, and D.T. Wong  {Clin. Cancer Res.} {\bf 14(19)} {6246-6252} {(2008)}.

\bibitem{InterlinkedInfo}
N.S. Gadewal, and S. M. Zingde {Bioinformation.} {\bf 6(4)} {169} {(2011)}.
  
\bibitem{BarabasiReview}
R. Albert, and A.-L. Barab\'asi, {Rev Mod Phys.} {\bf 74(1)} {47} {(2002)}.

\bibitem{BarabasiNat}
A.-L. Barabasi, and Z.N. Oltvai {Nat. Rev. Genet.} {\bf 5} {2} {(2004)}.

\bibitem{GohKI}
K.I. Goh, M. E. Cusick, D. Valle and et al. {Proc. Natl. Acad. Sci. U.S.A.} {\bf 104} {21} {(2007)}.

 \bibitem{individual_pro}
J.J. Hornberg, F.J. Bruggeman, H.V. Westerhoff, and J. Lankelma. {Biosystems.} {\bf 83(2)} {81-90} {(2006)}.
  
\bibitem{Mitra}
S. Mitra, S. Das, S. Das and S. Ghosal {Oral Oncol.} {\bf 48} {2} {(2012)}.

\bibitem{Wu}
C.H. Wu, R. Apweiler and A. Bairoch, {Nucleic Acids Res.} {\bf 34} {PMC1347523} {(2006)}.
   
\bibitem{spectra}
M. Molloy, and B. Reed, {Random Structures Algorithms.} {\bf 6} {161–179} {(1995)}. 
   
\bibitem{Newmann}
M. Newman, The structure and function of networks. Comput Phys Commun. {\bf 147} 40–45. (2002)

\bibitem{Reka}
R. Albert, H. Jeong, and A.L. Barabasi, {Nature.} {\bf 401} {130} {(1999)}.

\bibitem{Rivera}
M.T. Rivera, S.B. Soderstrom, and B. Uzzi, {Annu. Rev. Sociol.} {\bf 36} {91} {(2010)}.
   
\bibitem{Alok}
A. Yadav, and S. Jalan, Chaos. {\bf 25}, 043110 (2015).
  
\bibitem{Kitano}
H. Kitano, {Nat. Rev. Genet. and references therein.} {\bf 5(11)} {826} {(2004)}.

\bibitem{Ispolatov} 
I. Ispolatov, P.L. Krapivsky, A. Yuryev, Phys Rev E - Stat Nonlinear, Soft Matter Phys. {\bf 71} 1–22 (2005). 
   
\bibitem{Van}
P.V. Meighem, Graph Spectra for Complex Networks {Cambridge University Press, Cambridge.} {}{(2011)}{}.  
 
\bibitem{MatrixA}
H. Golub Gene, and F. Charles Van Loan. Matrix computations. JHU Press {\bf Vol. 3.} 2012.

\bibitem{Watts}
D.J. Watts, and S.H. Strogatz, {Nature.} {\bf 393} {440} {(1998)}. 
  
\bibitem{Signal}
L.M. Iakoucheva, C.J. Brown, J.D. Lawson, Z. Obradović, and A.K. Dunker, {Nature.} {\bf 323(3)} {573-584} {(2002)}.
    
\bibitem{Barabasi}
A.L. Barabasi, Linked {Plume Editors} {ISBN 0738206679} {(2002)}{}.
   

\bibitem{Clauset}
A. Clauset, C.R. Shalizi and M.E.J. Newman, {SIAM Review.} {\bf 4} {51} {(2009)}.
\bibitem{Hsing}
M. Hsing. {Nat. Biotechnol.} {\bf 30} {842} {(2000)}.   
\bibitem{pdb}
Berman, M. Helen, and et al. {Nucleic Acids Res.} {\bf 28(1)} {235} {(2000)}.  
\bibitem{Bio2006}
P.F. Jonsson, and P.A. Bates, {Bioinformatics.} {\bf 22} {18} {(2006)}.
\bibitem{Kessler}
B.M. Kessler, {Curr. Opin. Chem. Biol.} {\bf 17} {59} {(2013)}.
\bibitem{Haglund} 
K. Haglund, I. Dikic,  Protein Degrad. {\bf 4} 1–20  (2007).
\bibitem{Shpitzer}
Shpitzer, Thomas, G. Bahar, R. Feinmesser, and R.M. Nagler,  {J Cancer Res Clin Oncol.} {\bf 133} {613} {(2007)}.
\bibitem{Nagler}
R. Nagler, O. Ben-Izhak, D. Savulescu and et al. {Biophys Acta.} {\bf 1802.5} {454} {(2010)}. 
\bibitem{Mukherjee}
S. Mukherjee, and S.K. Das, {Curr. Mol. Med.} {\bf 12/4} {443} {(2012)}.
\bibitem{Force}
A. Force, M. Lynch, F.B. Pickett, A. Amores, Y.L. Yan, J. Postlethwait, {Genetics.} {\bf 151(4)} {1531} {(1999)}.
\bibitem{Conf}
M.E.J. Newman, S.H. Strogatz,  and D.J. Watts, {Phys. Rev. E.} {\bf 64} {026118} {(2001)}.
\bibitem{Aquir}
M. de Aguiar, and Y. Bar-Yam, {Phys.Rev.E.} {\bf 71} {016106} {(2005)}.
\bibitem{Zhang}
J. Zhang, {Trends Ecol. Evol.} {\bf 18(6)} {292} {(2003)}.  
\bibitem{Nowak}
M.A. Nowak, M.C. Boerlijst, J. Cooke, and J.M. Smith, {Nature.} {\bf 388(6638)} {167} {(1997)}. 
\bibitem{Frank}
S.A. Frank, {Curr. Bio.} {\bf 23(9)} {R343} {(2013)}.
\bibitem{Richardson}
S. Richardson, R. Sandra, C. Salvador‐Palomeque, and G.J. Faulkner, {BioEssays.} {\bf 36(5)} {475-481} {(2014)}.
\bibitem{Bailey}
J.A. Bailey, Z. Gu, R.A. Clark, K. Reinert, R.V. Samonte, S. Schwartz, and E.E. Eichler, {Science.} {\bf 297(5583)} {1003} {(2002)}.
\bibitem{Creixell}
P. Creixell, E.M. Schoof, J.T. Erler, and R. Linding {Nat. Biotechnol.} {\bf 30(9)} {842} {(2012)}.
\bibitem{Farkas_PRE_2001}
I.J. Farkas, I. Derenyi, A.-L. Barab\'asi, and T. Vicsek, {Phys. Rev. E.} {\bf 64} {026704} {(2001)}.
\bibitem{HubProt}
G. Kar, A. Gursoy, and O. Keskin, { PLoS Comput Biol.} {\bf 5(12)} {e1000601} {(2009)}.
\bibitem{NewmanPRE2003}
M.E.J. Newman, and J. Park,  {Phys. Rev. E.} {\bf 68} {036122} {(2003)}.
\bibitem{Newman_2004}
D. Lusseau, and M.E.J. Newman, {Proc. R. Soc. London B.} {\bf 271} {S477} {(2004)}.
\bibitem{Assortativity_Plos}
D.A. Pechenick, J.L. Payne, and J.H. Moore, {PLoS Comput Biol.} {\bf 10(8)} {e1003780} {(2014)}.
\bibitem{drug_resistance1}
M. Dean, T. Fojo, and S. Bates, {Nat Rev Cancer.} {\bf 5(4)} {275} {(2005)}.
\bibitem{drug_resistance2}
M.M. Gottesman, {Annu Rev Med.} {\bf 53(1)} {615} {(2002)}. 
\bibitem{TNM}
S. Patel, J. Shah, CA Cancer J Clin. {\bf 55} 242–258 (2005).
\bibitem{trait} P. Hansen, B. Floderus, K. Frederiksen, and C. Johansen, Cancer. {\bf 103(5)} 1082-1091 (2005).
\bibitem{tian}
Q. Tian, N. Price, and L. Hood, J Intern Med. {\bf 271} 111–121 (2012).
\end{thebibliography}
\end{document}